\documentclass[floatfix,aps,twocolumn,preprintnumbers,superscriptaddress,floatfix,prl]{revtex4-2}
\usepackage{bm}% bold math
\usepackage{lipsum}% http://ctan.org/pkg/lipsum
\usepackage{graphicx,amssymb}
\usepackage{dcolumn}% Align table columns on decimal point
\usepackage{bm}% bold math
\usepackage{xcolor}

\preprint{Lee \textit{et al.}}

\begin{document}

\title{Superconductivity reinforces charge-density-wave phase coherence across cuprates}

\author{H. Lee}\thanks{These authors contributed equally to this work.}\email[]{heemin@slac.stanford.edu}
\affiliation{Stanford Synchrotron Radiation Lightsource, SLAC National Accelerator Laboratory, Menlo Park, California 94025, USA}
\author{C.-T. Kuo}\thanks{These authors contributed equally to this work.}
\affiliation{Stanford Synchrotron Radiation Lightsource, SLAC National Accelerator Laboratory, Menlo Park, California 94025, USA}
\author{M. Fujita}
\affiliation{Institute for Materials Research, Tohoku University, Katahira 2-1-1, Sendai, 980-8577, Japan}
\author{C.-C. Kao}
\affiliation{Stanford Synchrotron Radiation Lightsource, SLAC National Accelerator Laboratory, Menlo Park, California 94025, USA}
\affiliation{SLAC National Accelerator Laboratory, Menlo Park, California 94025, USA}
\author{J.-S. Lee}\email[]{jslee@slac.stanford.edu}
\affiliation{Stanford Synchrotron Radiation Lightsource, SLAC National Accelerator Laboratory, Menlo Park, California 94025, USA}

\date{\today}% It is always \today, today,
             %  but any date may be explicitly specified

\begin{abstract}
For decades, superconductivity in high-$T_{\rm c}$ cuprates has been viewed as a competitor that suppresses charge-density-wave (CDW) order by reducing its amplitude and spatial extent. 
Here we show that this picture is incomplete, as superconductivity is accompanied by a systematic enhancement of CDW phase coherence across multiple cuprate families.
Using resonant soft x-ray scattering combined with a coherence-sensitive momentum-profile analysis, we uncover a BCS-like growth of phase coherence below $T_{\rm c}$, which phenomenologically manifests as the absence of CDW peak broadening and near-perfect wavevector locking. 
This enhancement remains visible even in a disorder-dominated regime created by long-term crystal aging and follows a common trend when compared with published data on Bi-, Hg-, Y-, and Nd-based cuprates. These results indicate that superconductivity reshapes CDW order in two distinct ways, suppressing its amplitude while strengthening its phase coherence, and reveal an additional phase-level interplay with lattice coupling in high-$T_{\rm c}$ cuprates.
\end{abstract}

\date{\today}
\maketitle

%%%%%%%%%%%%%%%%%%%%%%%%%%%%%%%%%%%%%%%%%%%%%%
%\section{INTRODUCTION}
%%%%%%%%%%%%%%%%%%%%%%%%%%%%%%%%%%%%%%%%%%%%%%
%
The relationship between charge density wave (CDW) order and superconductivity (SC) in cuprates has been cast almost entirely as a competition \cite{Berg1,Fradkin1,Ghiringhelli1,Chang1,Blackburn1,Blanco-Canosa1,Wu1,LeBoeuf1,Silva1,Lee1,Agterberg1,Wu2,Keimer1,Achkar1,Wen1,Huang1,Lee2,Arpaia1,Tabis3,Blanco-Canosa2,Croft1,Kivelson1,Mesaros1,Hayward1,Nie1,Campi1,Jang1}. When superconductivity is weakened by high magnetic fields, CDW correlations become stronger \cite{Chang1,Wu1}. Even high-field studies in YBa$_2$Cu$_3$O$_{6+y}$ that revealed field-induced CDW states with altered dimensionality, however, have not displaced this view \cite{Wu2,Wu3,Gerber1,Chang2}. Since most scattering works have relied on intensity-based (i.e., CDW volume) metrics \cite{Ghiringhelli1,Chang1,Blackburn1,Blanco-Canosa1,Silva1,Achkar1,Wen1,Huang1,Lee2,Arpaia1,Tabis3,Blanco-Canosa2,Croft1,Hayward1}, such approaches can overlook momentum-profile information, such as systematic changes in peak width or wavevector, that could potentially reveal hidden aspects of the SC-CDW orders’ electronic coupling and connect to signatures like phonon softening \cite{Lee2,Arpaia1,Tacon1,Kim1,Reznik1,Pintschovius1,Devereaux1,Chaix1} and phase-correlated optical dynamics \cite{Sugai1,Torchinsky1,Nguyen1,Wandel1,Jang2}. Here we show the opposite, that superconductivity reinforces, rather than suppresses, CDW phase coherence across diverse cuprate families.

%\textcolor{green}{A quantitative analysis of the momentum profile is challenging because the short-ranged and fluctuating nature of CDW order in cuprates leads to weak and broad ordering peaks} 
A quantitative analysis of the momentum profile is challenging due to the short-ranged and fluctuating nature of CDW order, which produces weak and broad peaks \cite{Kivelson1,Mesaros1,Hayward1,Nie1,Campi1,Jang1}. Extracting correlation length and phase coherence has been unreliable, especially below $T_{\rm c}$, due to noise, limits in low temperature sample environments, and scarce measurement time. Recent advances 
%\textcolor{green}{, for example,} 
in detecting efficiency \cite{Zhou1}, temperature-control 
%\textcolor{green}{led stages} 
\cite{Kuo1}, and 
%\textcolor{green}{optimized} 
acquisition strategies \cite{Wen1}, now make it possible to capture these weak features with the fidelity required for definitive analysis.

In this work, we apply a coherence-sensitive formalism 
%\textcolor{green}{(see Supplemental Material}
\cite{supp} that, for the first time, separates domain-size effects from intrinsic phase ordering, revealing a BCS-like growth of CDW phase rigidity below $T_{\rm c}$ with near-perfect wavevector locking. We demonstrate this effect in a prototypical La-based cuprate (La$_{2-x}$Sr$_x$CuO$_4$) and under a rare disorder-control condition 
%\textcolor{green}{(see Supplemental Material} 
\cite{supp}, mimicking conditions common to many cuprates \cite{Kivelson1,Mesaros1,Hayward1,Nie1,Campi1,Jang1}. Comparison with published data from Bi/Hg/Y-based cuprates \cite{Lee2,Arpaia1,Tabis3,Blanco-Canosa2,Croft1} confirms that this phase-coherence reinforcement is robust across structurally and electronically diverse cuprates. These results extend the long-standing competition paradigm by showing that superconductivity and CDW can be entangled not only as rivals but also as mutually reinforcing at the phase level \cite{NPhys2007,AdvMater2024}.

%======
%\subsection{1. Superconductivity sharpens CDW momentum structure}
In the prevailing competition-only picture, SC is expected to suppress CDW order below $T_{\rm c}$ in both amplitude and spatial correlation, producing a broader momentum profile. Any genuine narrowing would therefore be unexpected and would challenge this framework. Hints of such momentum-space evolution have appeared in scattering data on La-based cuprates \cite{Wen1,Croft1,Thampy1,Miao1,AdvMater2024} and other families \cite{Lee2,Arpaia1,Tabis3,Blanco-Canosa2}, but these have rarely been subjected to a full momentum-profile analysis. This is because the relevant CDW peaks are superimposed on strong resonant backgrounds, making precise extraction of their widths non-trivial even when signal-to-noise is adequate.

%%%%%%%%%%%%%%%%Fig1%%%%%%%%%%%%%%%%%%%%%%%%%%%%%%
\begin{figure}[t]
\begin{center}
\includegraphics[width=0.46\textwidth]{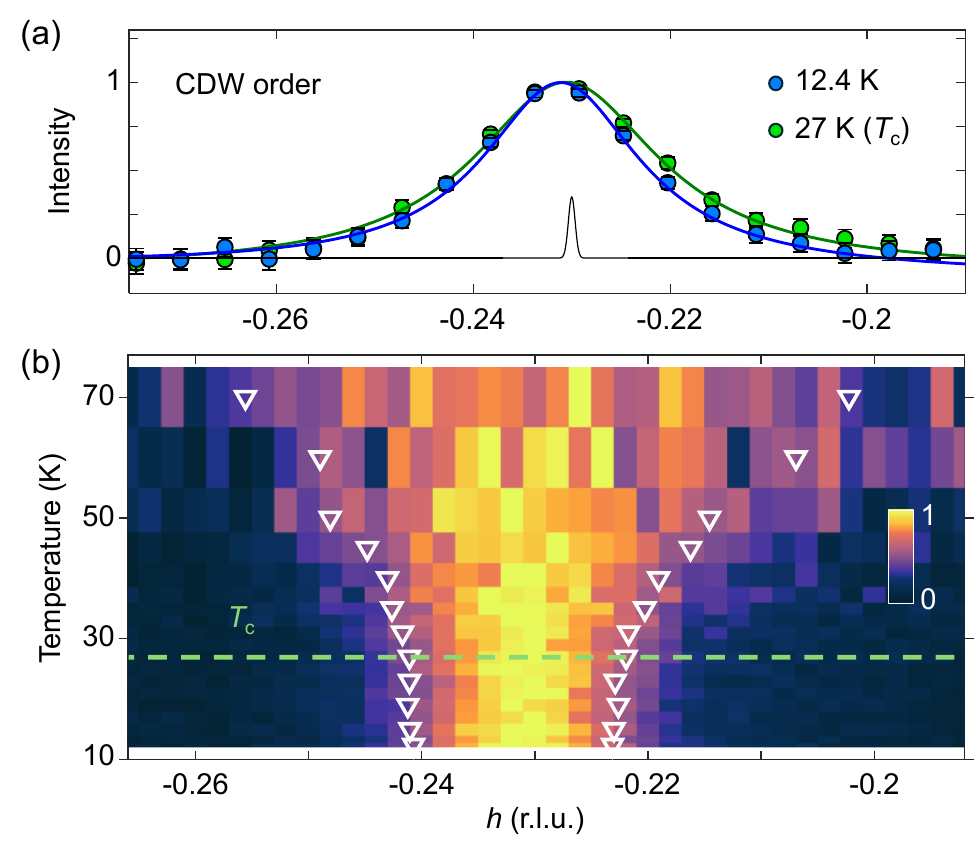}
\caption{Momentum profiles of CDW order across superconducting temperature. (a) CDW peak profiles along the $h$-direction measured at $T_{\rm c}$ = 27 K (green colored circles) and well below $T$ = 12.4 K (blue circles), together with corresponding single Lorentzian fits. The difference in peak width between two temperatures exceeds the instrumental resolution (black line) \cite{Kuo1}. (b) Two-dimensional map of the CDW profiles ($h$ vs. $T$). For each temperature, the profile is normalized to its maximum intensity. Inverted triangles mark the FWHM of each peak 
%\textcolor{green}{(see Supplemental Material for details} 
\cite{supp}.} \label{Fig1}
\end{center}
\end{figure}
%%%%%%%%%%%%%%%%%%%%%%%%%%%%%%%%%%%%%%%%%%%%%%

We address this question using high-statistics resonant soft x-ray scattering (RSXS) at Cu $L_{\rm 3}$-edge \cite{supp} on La$_{1.885}$Sr$_{0.115}$CuO$_4$ (LSCO), a composition with well-established stripe-like CDW correlations \cite{Wen1}. 
%
%High-quality LSCO single crystals were grown, and annealed to eliminate oxygen deficiencies. The superconducting transition temperature, $T_{\rm c}$ (27 K) was determined from the midpoint of the magnetic susceptibility transition. To achieve a fresh/clean $c$-axis-normal surface, each sample was resized (1.5 $\times$ 1.5 $\times$ 2.5 mm$^3$ in $a$ $\times$ $b$ $\times$ $c$ axes) and cleaved before delivering X-ray inside an ultra-high vacuum chamber (base pressure $\sim$ $10^{-10}$ Torr).
%
%All the RSXS were performed at beamline 13–3 of the Stanford Synchrotron Radiation Lightsource (SSRL, see Supplemental Material for details \cite{supp}). 
%
The high fidelity of our RSXS measurements \cite{Kuo1} allows us to track weak CDW peaks across the full temperature range. Figure 1(a) compares momentum profiles along $h$-direction from just above $T_{\rm c}$ (27 K) to deep in the superconducting state (12.4 K), showing a reduction of the full width at half maximum (FWHM --
%\textcolor{green}{hereafter denoted as}
$\Delta q_{\rm obs}$), from 0.023 ± 0.0005 to 0.018 ± 0.0005 r.l.u., corresponding to a 21 \% reduction that exceeds instrumental resolution, despite approximately 15 ± 0.2 \% drop in integrated intensity [Fig. S1 \cite{supp} and Fig. 3(c)]. Note that the effect is statistically significant and robust to variations in background modeling, line shape fitting, and measurement geometry.

To remove amplitude-related bias from domain-size effects, we normalize each profile to its maximum intensity [Fig. 1(b)]. The reduced peak width remains after normalization, with no evidence of the broadening expected from the competition picture. Importantly, the same behavior is independently reproduced in O $K$-edge RSXS measurements (Fig. S2 \cite {supp}), confirming that the effect is robust across different resonant channels. Such statistically robust behavior cannot be explained by resolution or domain-size effects alone, motivating the coherence-sensitive momentum-profile analysis introduced in the following section to isolate the intrinsic phase coherence of the CDW.

%======
%\subsection{2. Enhancement of CDW phase coherence below Tc}
The temperature evolution of the CDW peak width below $T_{\rm c}$ [Fig. 1(b)] was quantified by converting the measured FWHM (Fig. S1 \cite {supp}) into a real-space correlation length ($\xi$), providing a physically meaningful measure of the spatial extent of CDW correlations independent of amplitude changes [Fig. 2(a)]. At $T_{\rm c}$, $\xi$ $\approx$ 57.4 $\pm$ 0.5 \AA, corresponding to  $\sim$ 15 $a$-axis unit cells. At the lowest measured temperature (12.4 K), $\xi$ increases by roughly two unit cells. Although the total variation across the full temperature range is modest, its continuous growth through $T_{\rm c}$ indicates a clear change in CDW state concurrent with superconductivity.

Within the strict competition picture, the observed drop in integrated CDW intensity below $T_{\rm c}$ [Fig. S1 \cite {supp} and Fig. 3(c)] would normally be attributed to reduced domain size, which would also broaden the peak. To test this, we constructed an experimentally constrained scattering model incorporating the measured temperature dependencies of integrated intensity, 
%\textcolor{green}{momentum width} 
FWHM, and ordering vector. The resulting curve [red dashed line, Fig. 2(a)] reproduces the high-temperature behavior but fails to capture the continued growth of $\xi$ below $T_{\rm c}$, implying that an additional ordering mechanism emerges in the superconducting state.

%%%%%%%%%%%%%%%%Fig2%%%%%%%%%%%%%%%%%%%%%%%%%%%%%%
\begin{figure}[t]
\begin{center}
\includegraphics[width=0.49\textwidth]{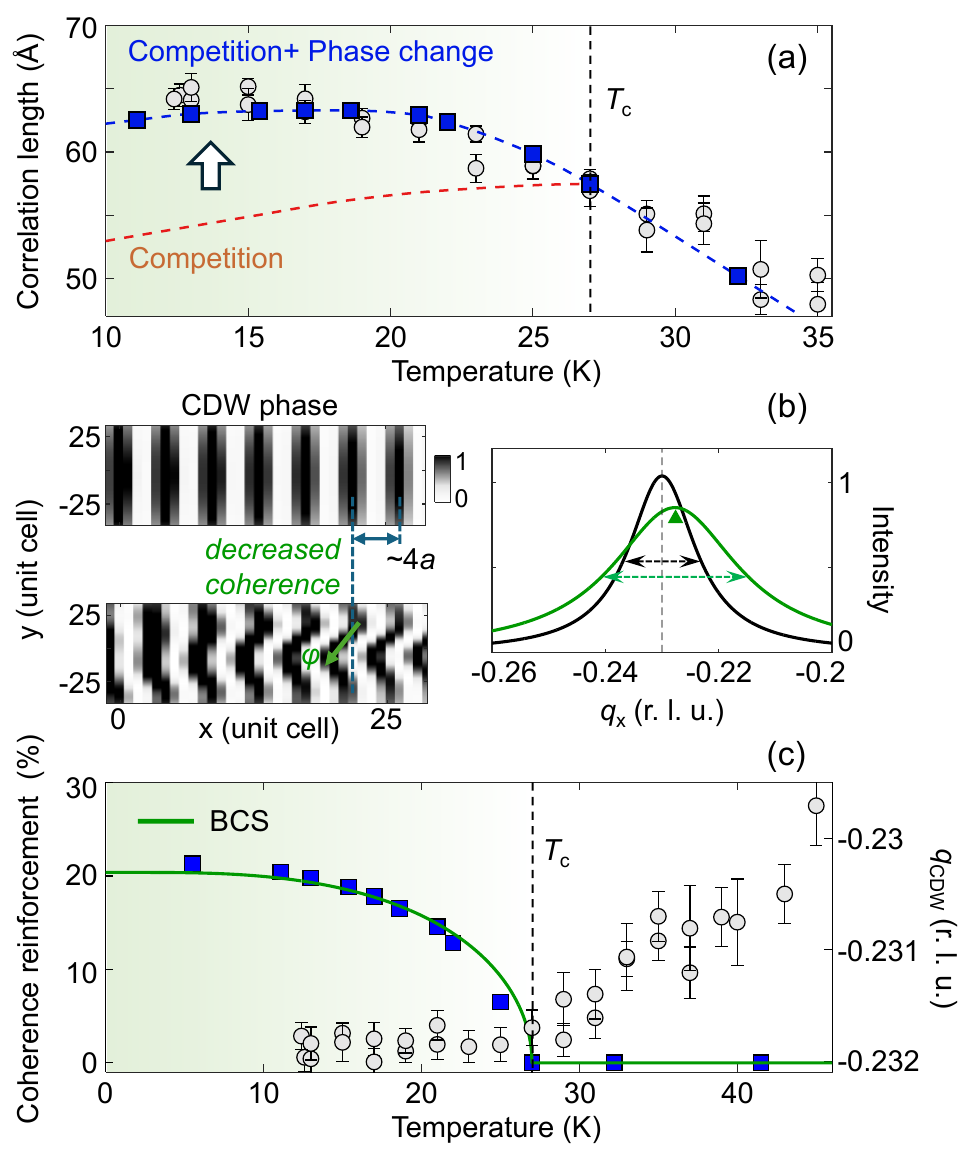}
\caption{Promotion of CDW phase coherence below $T_{\rm c}$. (a) Correlation length of CDW order (grey circles) as a function of temperature, obtained from the peak width using $\xi =$ 2/FWHM. Dashed curves show numerical simulations based on the scattering formalism with (blue) and without (red) a phase change below $T_{\rm c}$ 
%\textcolor{green}{(see Supplemental Material}
\cite{supp}. 
The blue squares are sampling points generated from the model curve, including points near the experimental temperatures.
The arrow denotes the change in CDW phase below $T_{\rm c}$. (b) Simulated CDW patterns for an ideally coherent phase (top) with $\sim 4a$ periodicity and for a decreased coherent (middle) with finite phase variations, $\varphi$. Bottom panel shows the corresponding calculated CDW peak profiles. 
This panel is an idealized sketch intended to highlight the qualitative effect of the phase-coherence channel on the CDW peak width and is not drawn from the experimental peak shapes shown in Fig. 1(a).
Arrows and the triangle indicate relative changes in FWHM and CDW wavevector, respectively. (c) Relative reinforcement of phase coherence (blue squares, left axis) below $T_{\rm c}$ extracted from $\xi$ in (a), fitted with a BCS-like gap function (green line). Measured CDW wavevectors (grey circles, right axis) are plotted versus temperature. Error bars represent one standard deviation (S.D.) of the fit parameters.} 
\label{Fig2}
\end{center}
\end{figure}
%%%%%%%%%%%%%%%%%%%%%%%%%%%%%%%%%%%%%%%%%%%%%%

To interpret the departure from the continuous growth of the CDW width through $T_c$, it is essential to recognize that the observed width $\Delta q_{\rm obs}(T)$ does not correspond to a single microscopic length scale. In standard CDW and smectic diffraction theory \cite{McMillan1976,TonerNelson1981,Zachar1998}, the width separates into two independent contributions 
%\textcolor{green}{(see Sec.~2 of the Supplemental Material} 
\cite{supp}:
\begin{equation}
\Delta q_{\rm obs}^{2}(T)
= \Delta q_{\rm dom}^{2}
+ \Delta q_{\varphi}^{2}(T),
\end{equation}
where $\Delta q_{\rm dom}$ represents disorder-dominated broadening that varies only weakly with temperature and limits the CDW domain size (and therefore governs the CDW intensity), while $\Delta q_{\varphi}(T)=\xi_{\varphi}^{-1}(T)$ reflects intrinsic phase coherence. The domain term captures the conventional competition between CDW order and superconductivity through the reduction of CDW volume below $T_{\rm c}$, whereas the phase term isolates the coherence channel that has not been accessible through intensity-based analyses.

When this decomposition is applied to the data, the temperature evolution of the observed width is governed primarily by the phase term. A decrease of $\Delta q_{\varphi}(T)$ below $T_{\rm c}$ signals enhanced CDW phase coherence. At low temperatures, where the domain contribution dominates, $\Delta q_{\rm obs}(T)$ may flatten or show a weak upturn even if coherence continues to improve; this behavior is a direct mathematical consequence of the two-channel relation and neither indicates a loss of coherence nor reflects a fitting artifact. In this interpretation, superconductivity suppresses the CDW amplitude while simultaneously reinforcing its phase coherence.

%\textcolor{green}{A natural framework for this behavior is provided by the liquid-crystalline description of stripe order} 
This behavior can be described within a liquid-crystalline framework of stripe order \cite{Kivelson1,McMillan1976,TonerNelson1981,Zachar1998}, where amplitude and phase fluctuations are decoupled. The local charge density along stripe $j$ may be written as
\[
\rho_j(x)
= \bar{\rho}
+ \rho_0 \cos\!\left[T_{\varphi}^j + L_{\varphi}(x)\right],
\]
where $\bar{\rho}$ and $\rho_0$ denote the average charge density and CDW amplitude, and $T_{\varphi}^j$ and $L_{\varphi}(x)$ describe transverse and longitudinal phase variations. This formulation smoothly interpolates between uniform-phase and reduced-coherence regimes, thereby separating genuine phase-ordering effects from domain-size broadening, as illustrated in Fig.~2(b). Reduced phase coherence broadens the momentum profile, whereas enhanced coherence suppresses peak broadening and stabilizes the ordering vector.

Using this coherence-sensitive formalism (Fig.~S3 \cite{supp}), we reproduce the observed growth of $\xi$ below $T_{\rm c}$ [blue dashed line in Fig.~2(a)] and quantify the enhancement of phase coherence [Fig.~2(c)], which increases by approximately $\Delta q_{\varphi}(T)\sim 20\%$ between $T_{\rm c}$ and 12.4~K. 
%
%This evolution follows a BCS-like temperature dependence, indicating a direct link to the superconducting condensate. 
%
This behavior can be phenomenologically parameterized using a BCS-like form, solely as a convenient descriptor of its rapid growth below $T_{\rm c}$. If only the domain contribution were present, the width would remain constant or broaden below $T_{\rm c}$, as shown by the red dashed line in Fig.~2(a). The observed behavior of $\Delta q_{\rm obs}(T<T_{\rm c})$ therefore requires an additional contribution, specifically a reduction of $\Delta q_{\varphi}(T)$, that is, an enhancement of intrinsic phase coherence.

Furthermore, the CDW wave vector ($q_{\rm CDW}$) nearly locks at $h \approx -0.2318 \pm 0.0005$ r.l.u.\ below $T_c$, consistent with coupling to the lattice and with phonon anomalies observed in RIXS at the superconducting onset \cite{Lee2,Arpaia1,Tacon1,Kim1,Reznik1,Pintschovius1,Devereaux1,Chaix1}. Equivalent measurements at the symmetry-related positive-$h$ position (Fig.~S4 \cite{supp}) confirm that this behavior is intrinsic and independent of scattering geometry. Together, these results reveal a regime in which superconductivity and CDW phase coherence evolve cooperatively, challenging the conventional competition paradigm.

%======
%\subsection{3. Persistence of phase coherence enhancement under disorder}
By separating the domain-limited and coherence-related contributions to the CDW peak width (Fig. 2), we next test whether this effect survives in the disorder-dominated regime characteristic of other cuprates \cite{Kivelson1,Mesaros1,Hayward1,Nie1,Campi1,Jang1}. In high-$T_{\rm c}$ cuprates, such as Bi- and Hg-based systems \cite{Lee2,Tabis3}, the CDW is inherently short-ranged and strongly influenced by quenched disorder \cite{Hayward1,Nie1,Jang1}, making it essential to determine whether the same coherence behavior can persist under such conditions.

We took advantage of a unique experimental opportunity in which the same single crystal from Figs. 1 and 2 had been kept under ambient laboratory conditions for over five years. Such prolonged aging introduces topological defects in the CuO$_2$ planes [Fig. 3(a)] within the soft x-ray probing depth, while preserving the bulk composition and growth history. 
%\textcolor{green}{Importantly,} 
The superconducting transition temperature remains unchanged within experimental uncertainty (%\textcolor{green}{see} 
Fig.~S6 \cite{supp}).
This before–after comparison within an identical specimen removes sample-to-sample variability and provides a controlled platform for studying disorder effects (see Sec.3 of the Supplementary 
%\textcolor{green}{Material for the characterization} 
\cite{supp}).

%%%%%%%%%%%%%%%%Fig3%%%%%%%%%%%%%%%%%%%%%%%%%%%%%%
\begin{figure}[t]
\begin{center}
\includegraphics[width=0.49\textwidth]{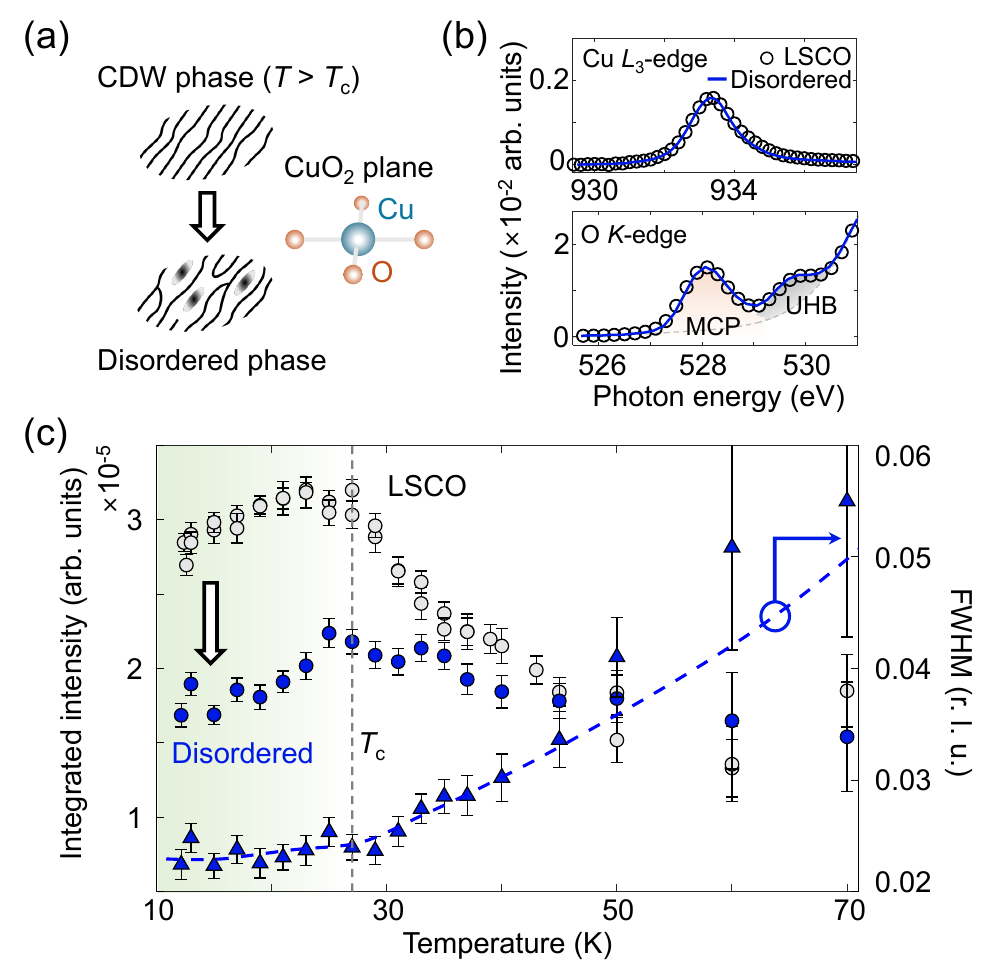}
\caption{CDW phase coherence under the disorder. (a) Schematic cartoons of CDW order in the normal state (i.e., $T > T_{\rm c}$) and in disordered state, illustrating the charge density topology in CuO$_2$ plane. (b) Cu $L_3$-edge and O $K$-edge XAS spectra of LSCO (circles) and its relatively disordered LSCO (blue lines). For O XAS spectra, red and gray shaded regions indicate the fitted MCP (or so-called Zhang–Rice singlet state) and the upper Hubbard band (UHB), respectively. Dashed line marks the background level. (c) Temperature dependence of the integrated CDW 
%\textcolor{green}{peak} 
intensity (i.e., CDW volume) for the disordered LSCO (blue circles) compared with the pristine one (grey circles). The arrow highlights the reduced CDW volume in the disordered crystal. The right axis presents the corresponding 
%\textcolor{green}{momentum width (} 
FWHM of the disordered LSCO (
%\textcolor{green}{blue} 
triangles) and analyze (blue dashed line) through the coherence-sensitive formalism (see Supplemental Material \cite{supp}).  All error bars represent 1 S.D. of the fit parameters.}
\label{Fig3}
\end{center}
\end{figure}
%%%%%%%%%%%%%%%%%%%%%%%%%%%%%%%%%%%%%%%%%%%%%%

Cu $L_{\rm 3}$-edge and O $K$-edge XAS spectra [Fig. 3(b)] show no detectable difference between the pristine and aged states. 
%
%\textcolor{green}{In particular,} 
The ratio \cite{Chen1,Abbamonte1,Abbamonte2,Peets1} between the so-called mobile carrier peak (MCP) or hole state ($\sim$ 528.1 eV) and the UHB 
%\textcolor{green}{upper Hubbard band} 
($\sim$ 529.7 eV) remains unchanged, indicating that the carrier concentration is preserved and that disorder primarily weakens translational CDW order without altering the electronic ground state. At the same time, independent measurements (Figs. S7 and S9) confirm that the aged LSCO crystal is not strictly identical to the pristine sample. While the bulk-sensitive TFY spectra remain essentially unchanged, the surface-sensitive TEY signal and subtle changes in the Bragg scattering profile indicate weak structural modifications associated with long-term aging.

RSXS measurements on the aged crystal (Fig. 3(c) and Fig. S5 \cite{supp}) reveal a $\sim$ 60 \% reduction in CDW volume and a clear broadening of its momentum width, consistent with a transition to a weak, short-range ordered state. Importantly, when analyzed using the same coherence-sensitive scattering formalism as in Figs. 1 and 2, the residual CDW exhibits no broadening of the momentum width below $T_{\rm c}$, indicating that the superconductivity-induced phase coherence enhancement remains operative despite the presence of strong disorder (Fig. S5(e) \cite{supp}). These results place the aged crystal in the same disorder-dominated regime as many other cuprate families, establishing the basis for direct comparison of SC-CDW interplay under comparable conditions.

%%%%%%%%%%%%%%%%Fig4%%%%%%%%%%%%%%%%%%%%%%%%%%%%%%
\begin{figure}[t]
\begin{center}
\includegraphics[width=0.47\textwidth]{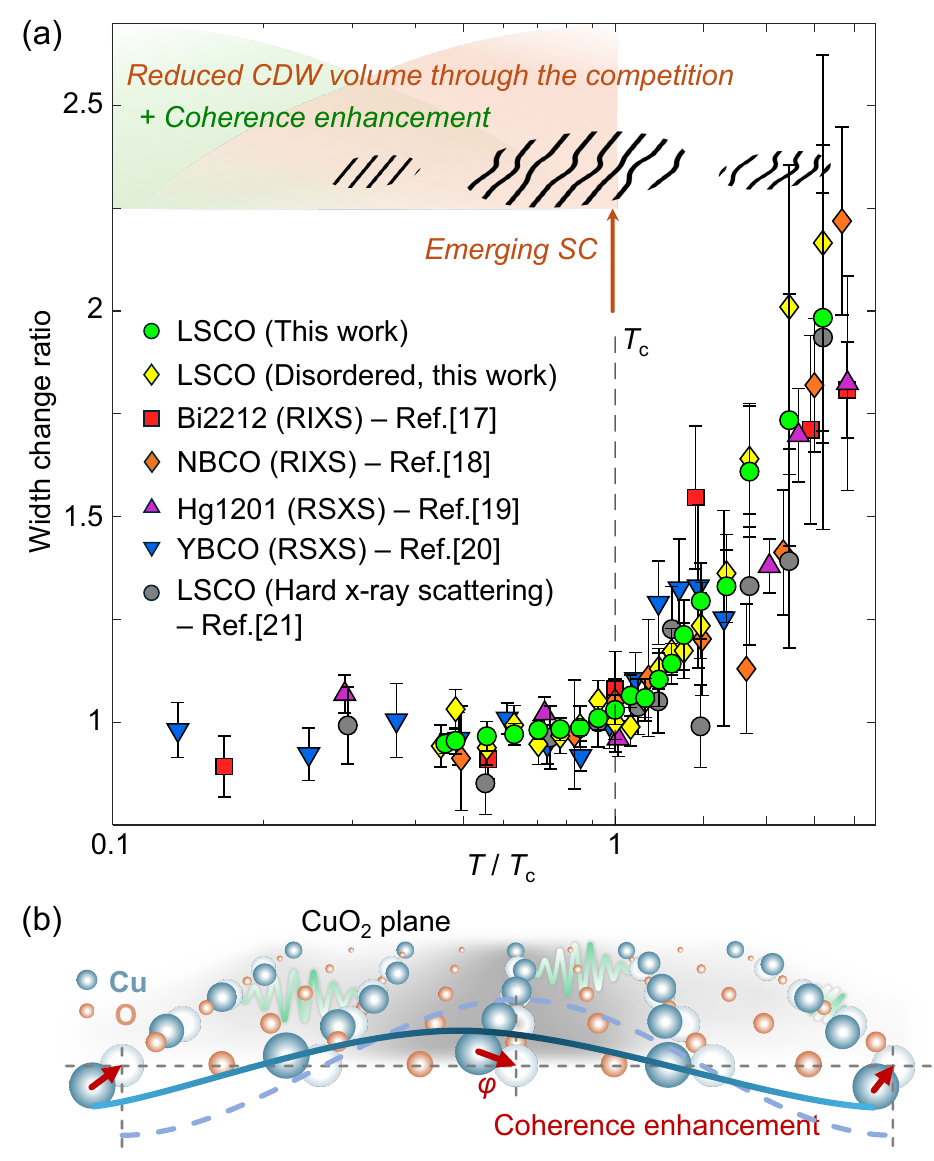}
\caption{CDW-SC interplay through phase-coherence reinforcement across cuprate families. (a) Temperature dependence of the CDW peak-width change ratio, normalized to the width at $T_{\rm c}$, for various high-$T_{\rm c}$ cuprate families measured by RSXS, RIXS, and hard x-ray scattering. Data include this study, LSCO, disordered LSCO, and published data from Bi2212 [17], NBCO [18], Hg1201 [19], YBCO [20] (blue inverted triangles), and LSCO [21] from hard x-ray scattering. The top schematic illustrates an expanded framework for the SC-CDW relationship, where superconductivity can reinforce CDW phase coherence in addition to competing with its volume. (b) Schematic cartoon illustrating the variation in the CDW phase within the liquid-crystalline character of the superconducting CuO$_2$ plane. Light and dark blue shading distinguish the more or less coherent CDW phases, respectively. Red arrows indicate changes in the CDW phase coherence, expressed as variation in $\varphi$, potentially driven by collective excitations such as phonon activity (green oscillations).}
\label{Fig4}
\end{center}
\end{figure}
%%%%%%%%%%%%%%%%%%%%%%%%%%%%%%%%%%%%%%%%%%%%%%

%======
%\subsection{4. Phase-coherence across diverse cuprates} 
%\textcolor{green}{To test whether the phase coherence enhancement identified in LSCO is unique or generic} 
To test whether the phase coherence enhancement in LSCO is generic, we compile in Fig. 4(a) the normalized CDW momentum width from this work (pristine and disorder dominated LSCO) together with published data on Bi-\cite{Lee2}, Hg-\cite{Tabis3}, Y-\cite{Blanco-Canosa2}, and Nd-based \cite{Arpaia1} cuprates from RSXS, RIXS, and non-resonant hard x-ray scattering. We note that CDW (or CO) phenomena in different cuprate families may arise from distinct microscopic mechanisms, and the comparison made here therefore focuses only on the phenomenological behavior of the CDW peak width. 
%\textcolor{green}{Remarkably,} 
%Despite wide differences in composition, CDW dimensionality, and technique, the temperature evolution of the width collapses onto a nearly universal curve, showing no broadening below $T$/$T_{\rm c}$ $\approx$ 1. 
%
Despite differences in composition, dimensionality, and technique, the temperature dependence of the CDW peak width shows a nearly universal behavior.
%
%This indicates that superconductivity induced growth of CDW phase coherence is a robust property of the cuprate phase diagram.
%
This indicates that superconductivity-induced CDW phase coherence is a robust feature across cuprates.

As sketched in Fig. 4(a) (top), the conventional SC-CDW competition picture remains valid, since CDW volume suppression below $T_{\rm c}$ is ubiquitous \cite{Berg1,Fradkin1,Ghiringhelli1,Chang1,Blackburn1,Blanco-Canosa1,Wu1,LeBoeuf1,Silva1,Lee1,Agterberg1,Wu2,Keimer1,Achkar1,Wen1,Huang1,Lee2,Arpaia1,Tabis3,Blanco-Canosa2,Croft1,Kivelson1,Mesaros1,Hayward1,Nie1,Campi1,Jang1}, but our results establish that it is accompanied by concurrent phase coherence enhancement. 
%
%Figure 4(b) illustrates 
%%\textcolor{green}{one possible} 
%a scenario in which entering the superconducting state allows theoretically predicted liquid crystal-like CDW domains \cite{Kivelson1,McMillan1976,TonerNelson1981,Zachar1998} to partially rephase, potentially mediated by lattice degrees of freedom and electron phonon coupling \cite{NPhys2007}. 
%
Figure 4(b) illustrates a scenario in which entering the superconducting state allows partial rephasing of CDW domains, potentially mediated by lattice coupling \cite{Kivelson1,McMillan1976,TonerNelson1981,Zachar1998,NPhys2007}.
This cooperative framework naturally reconciles several previously disconnected observations, including phonon softening in the CDW state seen by RIXS/IXS \cite{Lee2,Arpaia1,Tacon1,Kim1,Reznik1,Pintschovius1,Devereaux1,Chaix1} and optical signatures of phase-sensitive CDW dynamics across $T_{\rm c}$ \cite{Sugai1,Torchinsky1,Nguyen1,Wandel1,Jang2}. In this view, superconductivity reshapes the low-energy susceptibility and effectively renormalizes the CDW phase stiffness \cite{Arpaia1,Tacon1,Reznik2}, so that when it is suppressed the latent tendency toward CDW coherence becomes more apparent. Such a perspective may also help to explain the emergence of long-ranged three-dimensional CDW characteristics \cite{Wu3,Gerber1,Chang2} in YBCO when superconductivity is suppressed by either high magnetic fields or photoexcitation.

Finally, we note that superconductivity-enhanced CDW phase coherence may be viewed in the broader context of intertwined spin, charge, and pair-density-wave 
%\textcolor{green}{(PDW)} 
orders in cuprates. Recent work on LSCO \cite{arxivLSCO} has reported an enhancement of the spin-stripe correlation length and discussed its possible relation to PDW physics, suggesting that different components of an intertwined stripe order may share, at least partially, a common phase stiffness. Earlier studies have likewise questioned the notion of purely competitive CDW–SC interplay \cite{NPhys2007,AdvMater2024}. While our experiment directly probes only the charge sector, the coherence reinforcement observed here is naturally compatible with these broader intertwined-order frameworks and motivates future studies that jointly probe spin-, charge-, and pair-order correlations.

%======
%\section{Discussion}
In summary, our results reveal an unrecognized yet fundamental component of the SC-CDW relationship in cuprates, where superconductivity can enhance CDW phase coherence even while competing with its amplitude. This observation moves beyond long-standing debates about phase separation and instead points to a deeper electronic correlation between the two orders. Such a link suggests that superconductivity and CDW order may be coupled manifestations of a shared underlying electronic state, offering a basis for a more unified understanding of high-$T_{\rm c}$ superconductivity. By reframing the interplay from a competition-only picture to one that also includes phase-level reinforcement, our work further provides a conceptual path for reconciling diverse experimental observations across materials and techniques. The persistence of this effect in disorder-dominated regimes raises key questions about its microscopic origin and role in the superconducting mechanism. Addressing these questions will require targeted studies using collective modes probes, high magnetic fields, and optical tuning to directly test the dynamical and symmetry aspects of the coupling. More broadly, these findings encourage a shift from viewing intertwined orders solely as competitors to recognizing their potential for mutual reinforcement at the phase level.

%%%%%%%%%%%%%%%%%%%%%%%%%%%%%%%%%%%%%%%%%%%%%%
%\section*{ACKNOWLEDGMENTS}
%%%%%%%%%%%%%%%%%%%%%%%%%%%%%%%%%%%%%%%%%%%%%%

$Acknowledgments$—We thank Steven A. Kivelson and Vivek Thampy for insightful discussions. All X-ray experiments were carried out at beamline 13-3 in SSRL, SLAC National Accelerator Laboratory, supported by the U.S. Department of Energy, Office of Science, Office of Basic Energy Sciences under Contract No. DE-AC02-76SF00515. M.F. was supported by Grant-in-Aid for Scientific Research(S) (GrantNo.21H04987).
%
%%%%%%%%%%%%%%%%%%%%%%%%%%%%%%%%%%%%%%%%%%%%%%
% Create the reference section using BibTeX:
%%%%%%%%%%%%%%%%%%%%%%%%%%%%%%%%%%%%%%%%%%%%%%
% \bibliography{LSCO_4th}% 
%apsrev4-2.bst 2019-01-14 (MD) hand-edited version of apsrev4-1.bst
%Control: key (0)
%Control: author (8) initials jnrlst
%Control: editor formatted (1) identically to author
%Control: production of article title (0) allowed
%Control: page (0) single
%Control: year (1) truncated
%Control: production of eprint (0) enabled
%

\end{document}